\documentstyle[12pt]{article}
\renewcommand{\baselinestretch}{1.2}

\def\Tr{\mathop{\mbox{Tr}}\,}

\def\di{\mbox{d}}
\newcommand{\bea}{\begin{eqnarray}}
\newcommand{\be}{\begin{equation}}
\newcommand{\eea}{\end{eqnarray}}
\newcommand{\ee}{\end{equation}}
\newcommand{\nn}{\nonumber}
\newcommand{\spav}[1]{\parbox{1mm}{\vspace*{#1}}}

\begin{document}

\begin{titlepage}
\spav{3cm}

\begin{center}
{\Large\bf About the Matrix model computation}\\
{\Large\bf of three-graviton scattering}\\ 
\spav{1.8cm}\\
{\large Marco Fabbrichesi, Gabriele Ferretti and Roberto Iengo}
\spav{1cm}\\
{\em Istituto Nazionale di Fisica Nucleare, Sezione di Trieste}\\
{\em Scuola Internazionale Superiore di Studi Avanzati (SISSA)\\
Via Beirut 2-4, I-34013 Trieste, Italy} \\
\end{center}

\noindent
\hrulefill

{\small 
\noindent 
{\sc \bf Abstract}\\
\noindent
It is briefly explained why recent claims about the vanishing of
the one-loop effective potential in Matrix theory, thus invalidating the
possible agreement with supergravity, do not hold.}

\noindent
\hrulefill

\vfill

\flushleft{June 1998}

\end{titlepage}

\newpage 
\setcounter{footnote}{0}
\setcounter{page}{1}

In our paper~\cite{FFI}, we argued that it is possible for finite-$N$
Matrix theory to reproduce the supergravity amplitude for three-graviton
scattering, contrary to what previously claimed in~\cite{DR,D2}. Our
prediction is now fully borne out by the computation of~\cite{OY}, where it is
shown that there is perfect agreement between the two theories, numerical
coefficients included. In the meantime, two papers~\cite{WT,EG}
 came out, based
on an effective action approach, arguing that such agreement is impossible.
In particular, the two papers disputed our result about the non-vanishing
of the relevant one-loop effective potential which we
 used in discussing the agreement 
between Matrix theory and supergravity. This short note
addresses these objections and show why they do not hold.

Matrix theory is defined by a
 supersymmetric $SU(N)$ Yang-Mills (YM) field theory in
$1+0$ dimensions with the classical action
\be
S (A_\mu,\Psi) = 
\int \di t\: \left( - \frac{1}{2} \Tr F^{\mu\nu} F_{\mu\nu} +i \Tr \bar \Psi 
\Gamma \cdot D \Psi \right) \, ,
\ee
where $ F_{0k} = \partial_t A_k + i[A_0,A_k]$, $F_{jk} = i[A_j,A_k]$
and $D_0\Psi =  \partial_t \Psi + i[A_0, \Psi]$, $D_k\Psi =  i[A_k, \Psi]$.
Throughout this note, Greek indices go from 0 to 9 and Latin ones from 1 to 9.

In order to compare  the three-graviton scattering
 in supergravity with finite-$N$ Matrix theory, one has to
make a two-loop computation for $SU(3)$ YM
 in the presence of the following classical external field
\be
   B_k =\pmatrix{0 & 0     & 0     \cr
                 0     & r_2 - r_1 & 0     \cr
                 0     & 0     & r_3-r_1 \cr}_k \, , \label{B}
\ee 
which is solution
of the classical equations of motion when the graviton positions
${\bf r}_{1,2,3}$ depend linearly  on time.
We can fix the center of mass by writing
${\bf r}_1+{\bf r}_2+{\bf r}_3 =0$. We assume that
$|{\bf r}_2-{\bf r}_1|$ and $|{\bf r}_3-{\bf r}_1|$ are much larger than
$|{\bf r}_2-{\bf r}_3|$.
Notice that $B_k$ contains an irrelevant decoupled $U(1)$ component that we
retain for convenience.

We quantize by setting
\be
A_\mu = B_\mu + A^f_\mu \, ,
\ee
where $B_\mu = (0, B_k)$ and $A^f_\mu$ is the quantum (fluctuating) gauge
field.

We use the gauge fixing defined by
\be
D^\mu(B)\, A^f_\mu = \partial_t A^f_0 + i [B_k,\,A^f_k] \, .
\ee

We have to perform the functional integration
\be
e^{-F(B)} = \int [\di A^f_\mu] [\di \Psi] [\di \bar C] [\di C]
e^{-S( B_\mu + A^f_\mu, \Psi) - \int \left[D^\mu(B)\, A^f_\mu \right]^2 -
\int \bar C D^\mu(B) D_\mu(A) C}     \, .
\ee
The  2-loop action $F_{(2)}(B)$ is to be compared with the supergravity
amplitude for the three-graviton scattering.

We write
\be
A_\mu = \pmatrix{a_1 & a_2^\dag     &  a_3^\dag      \cr
                  a_2     & Y_{22} &   Y_{23}    \cr
                   a_3    &   Y_{32}    &  Y_{33} \cr}_\mu \, ,
\ee
where the $2\times 2$ matrix $Y_\mu = B_\mu + Y^f_\mu$ (by a slight abuse
of notation, we also denote by $B_\mu$ the appropriate $2\times 2$ sub-matrix
of (\ref{B})); $a_{1,2,3}$ and
$Y^f$ are quantum (fluctuating) fields.

For our choice of background~(\ref{B}), $a_{2\mu}, a_{3\mu}$ are
the {\it heavy} modes since their mass terms are proportional to the large
distances $|{\bf r}_2-{\bf r}_1|$ and $|{\bf r}_3-{\bf r}_1|$.
Similarly, for the fermions and the ghosts, we denote by 
$\psi_{2}, \psi_{3}, c_2, c_3, \bar c_2, \bar c_3$ the respective 
heavy modes, and by $\phi$, ${\cal C}$,  $\bar {\cal C}$ the remaining
{\it light} fields. 

We can then perform the functional integration in two steps. First, we
compute at 1-loop
\bea
&&e^{-W_{(1)}(Y^f, a_1, \phi, \bar {\cal C}, {\cal C}; B)}=   \\
&&\int [\di a_2 \di a_3] [\di \psi_2 \di\psi_3] [\di \bar c_2 \di\bar c_3] 
[\di c_2 \di c_3]
e^{-S( B_\mu + A^f_\mu, \Psi) - \int \left[ D^\mu(B)\, A^f_\mu \right]^2 -
\int \bar C D^\mu(B) D_\mu(A) C} \, . \nn 
\eea

To complete the computation one has to integrate over the light
modes:
\be
e^{-F(B)} = \int [\di Y^f] [\di a_1] [\di \phi] 
[\di \bar {\cal C}] [\di  {\cal C}]
e^{-W_{(1)}(Y^f, a_1, \phi, \bar {\cal C}, {\cal C}; B)} \, .     
\label{eqA}
\ee

Therefore, one has to compute 
\bea
W_{(1)} & = &V_{(1)} (Y^f; B) \\
& &  + \;  \mbox{terms with $a_1$}\; + \;
\mbox{derivative terms}\;  +  \;
\mbox{fermionic and ghost terms} \, , \nn  \label{eq10}
\eea
where $V_{(1)}$ is the potential---that is the term with no derivatives---of
the field $Y^f$.

 In our paper~\cite{FFI}, we have found that $V_{(1)} (Y^f; B) \neq 0$ 
for generic $Y^f$. In fact, let us take
\be
Y^f_k = \pmatrix{ 0 & x   \cr
                   x^{\dag}     &   0   \cr}_k \, ,
\ee
where $x_k$ is the gauge field component 
in the $k$ direction having the light mass
term $|{\bf r}_2-{\bf r}_3|$ (it is called $x^{\alpha^1}_k$ in~\cite{FFI}).
We obtain that the part of $V_{(1)}$ quadratic
in the field ${\bf x}$ is
\be
 V_{(1)} (Y^f; B) = \frac{|{\bf R}_1 \cdot {\bf x}|^2
-|{\bf R}_1|^2 |{\bf x}|^2}{R_2 R_3 (R_2 + R_3)} \, , \label{eqB}
\ee
where ${\bf R}_1= {\bf r}_2-{\bf r}_3$ and cyclic.

In the same paper, we have shown that by performing the 
functional integration over ${\bf x}$ and using eq.~(\ref{eqB}) 
as its potential, we obtain a term for $F_{(2)}(B)$ which has the same
behavior as the supergravity amplitude.

The papers~\cite{WT} and \cite{EG} challenge our result, by claiming that
$ V_{(1)} = 0$. However, both papers deal with a potential $V_{(1)}$ for
field configurations that are not the relevant ones.

In the appendix of ref.~\cite{WT}, it is shown
 that $ V_{(1)}(0;Y) = 0$ for any $Y$.
We have checked this result and found it to be true. This is however irrelevant
for the computation of eq.~(\ref{eqA}), which is the one  needed for the
comparison with supergravity. In fact, in eq.~(\ref{eqA}) one must integrate
$e^{-W_{(1)}}$ (which contains $V_{(1)}$, 
see eq.~(\ref{eq10})) over the fluctuations $Y^f$ keeping $B$ fixed. 

In ref.~\cite{EG} the computation is done
for the case when ${\bf R}_1$ and ${\bf x}$ are
parallel (see eqs.~(3) and (34) of ref.~\cite{EG}). In this
case, $V_{(1)} = 0$, which is true, as it can be seen from 
eq.~(\ref{eqB}). However, this is just a special choice
of ${\bf x}$ whereas in eq.~(\ref{eqA})
one must integrate over all the possible  ${\bf x}$.

%
\renewcommand{\baselinestretch}{1}

\end{document}